\documentclass{aa}

\usepackage{psfig,graphics}
\begin{document}

   \thesaurus{03 
              (11.08.1;  
               11.19.2; 
               11.19.3; 
               11.09.4; 
               11.19.6)} 
   \title{Extraplanar diffuse ionized gas in a small sample of nearby edge--on 
galaxies
   \thanks{Based on observations collected at the European Southern 
        Observatory, La Silla, Chile}}

   \author{J.~Rossa
          \inst{}
          \and
          R.--J.~Dettmar
          \inst{}}
	 
   \offprints{jrossa@astro.ruhr-uni-bochum.de}

    \institute{Astronomisches Institut der Ruhr--Universit\"at Bochum, 
              D--44780 Bochum, Germany}

   \date{Received 21. October 1999 / Accepted 20. April 2000}

   \maketitle

   \begin{abstract}

We present narrowband H$\alpha$ imaging data of a small survey of nearby 
edge--on spiral galaxies, aiming at the detection of `extraplanar' diffuse 
ionized gas (DIG). A few of our studied edge--on spirals show signs of 
disk--halo interaction (DHI), where extended line emission far above the 
galactic plane of these galaxies is detected. In some cases an extraplanar 
diffuse ionized gas (eDIG) layer is discovered, e.g.,\, \object{NGC4634}, 
\object{NGC\,3044}, while other galaxies show only filamentary features
reaching into the halo (e.g.,\,\object{IC\,2531}) and some galaxies show no 
sign of eDIG at all. The extraplanar distances of the DIG layer in our 
narrowband H$\alpha$ images reach values of $z\leq$ 2\,kpc above the galactic 
plane. The derived star formation rates ($SFRs$) from the H$\alpha$ flux of 
the studied galaxies range from $0.05-0.7\,\rm{M_{\sun}\,yr^{-1}}$, neglecting 
a correction for internal absorption. The variation of the $SFR$ values among 
our sample galaxies reflects the diversity of star formation within this 
sample. A diagnostic diagram is introduced, which allows to predict the 
existence of gas halos in `quiescent' galaxies based on the ratio $\rm{S_{60}
/S_{100}}$ versus $L_{\rm{FIR}} / D^2_{25}$ in this diagram. We compare the 
positions of the non--starburst galaxies with starburst galaxies, since these 
galaxies populate distinct positions in these diagrams.

   \end{abstract}

%

\section{Introduction}

In recent years diffuse ionized gas (DIG) frequently also called warm ionized 
medium (WIM) has been identified as an important component of the ISM, in 
particular with regard to the influence of SF on the large scale distribution 
and physical properties of the ISM. This gas component typically has a very 
low electron density of $</n/>\,\sim 0.08\,\rm{cm^{-3}}$ (in the disk) which 
decreases exponentially towards the halo and is characterized through a 
temperature of $T=$ 8000--10000\,K. For a detailed review on recent 
developments concerning the disk--halo connection, which is briefly 
described below, we refer to Dettmar (\cite{De95}) or Dahlem (\cite{Da97b}). 

In our own galaxy DIG was first detected as an extraplanar gas layer (Hoyle 
\& Ellis \cite{HoEl63}) by radio observations. In the early seventies 
H$\alpha$ observations were performed which also showed an extraplanar layer, 
(see e.g.,\,Reynolds \cite{Re84}) which is now known as the `Reynolds layer'. 
Only as recently as 1990 this gas component has been detected in external 
galaxies outside traditional \ion{H}{ii} regions (Dettmar \cite{De90}; 
Rand et al. \cite{Ra90}) 

Since the DIG is traced by H$\alpha$ emission, several studies have been 
dedicated to detect eDIG in external galaxies with the use of narrowband 
H$\alpha$ CCD imaging, preferentially in edge--on galaxies, where the halo 
separates from the disk (e.g.,\,Pildis et al. \cite{Pi94}; Rand et al. 
\cite{Ra92}; Rand \cite{Ra96}; this work). About two dozen galaxies have been 
detected up to now that show signs of disk--halo interaction (DHI). 
Subsequent longslit spectroscopy has been performed for a few galaxies 
including NGC\,891 (Dettmar \& Schulz \cite{DeSc}; Keppel et al. \cite{Ke91}; 
Rand \cite{Ra97}; Rand \cite{Ra98}), NGC\,4631 (Golla et al. \cite{GoDo}), 
NGC\,2188 (Domg\"orgen \& Dettmar \cite{DoDe}), NGC\,1963 \& NGC\,3044 
(T\"ullmann \& Dettmar \cite{TuDe}), among a few others. DIG detections in 
starburst galaxies seem to be a common feature, as it was evidenced by an 
investigation by Lehnert \& Heckman (\cite{LeHe}). DIG typically reaches 
scale--heights in edge--on galaxies of $\sim$1--2\,kpc, but as in the case of 
NGC\,891 spectroscopic investigations have shown that DIG even can be 
detected at extraplanar distances of up to 5\,kpc (Rand \cite{Ra97}).   

The most likely process for ionizing the DIG is photoionization (Mathis 
\cite{Ma86}; Domg\"orgen \& Mathis \cite{DoMa}). Although photoionization by 
OB stars (e.g.,\,Miller \& Cox \cite{MiCo}; Dove \& Shull \cite{DoSh}) is 
regarded as the primary process, other mechanisms have been invoked 
such as shock-ionization (Chevalier \& Clegg \cite{ChCl}), and turbulent 
mixing layers (Slavin et al. \cite{SlSh}) to account for the observed 
emission line ratios. 

A mechanism for the transport of gas and radiation into the halo has been 
formulated in the late eighties (Norman \& Ikeuchi \cite{NoIk}). The gas 
emanates from star forming regions in the disk of the galaxies into the halo 
via so called {\em chimneys}. This is a modified theoretical description of 
the formerly developed theory of galactic fountains (Shapiro \& Field 
\cite{ShFi}). In the chimney scenario gas is driven by collective supernovae. 
Starburst driven winds that cause outflows may also play an important role, 
at least in nuclear starburst galaxies (Heckman et al. \cite{HAM}). 

A larger sample of starburst galaxies has been studied by Lehnert \& Heckman 
(\cite{LeHe}). Since the DIG is generally believed to be correlated with the 
star formation activity in the underlying galaxy disk, in starburst galaxies 
DIG is detected relatively frequently, and seems to be a common feature, 
wherby in normal galaxies not all show any disk--halo interaction. A minimum 
energy input to the ISM is obviously necessary in order to show any outflow 
phenomena. Therefore there is the demand to study more edge--on galaxies in 
order to make quantitative statements. This first mini--survey, which we 
present in the following chapters, is the first part of a larger and much more 
quantitative survey which is currently under investigation (Rossa \& Dettmar, 
in prep.)   

While it is spectroscopically possible to obtain physical parameters using 
diagnostic line ratios (Osterbrock \cite{Os89}), narrowband imaging can be 
used to investigate the morphology of the eDIG in detail and allows 
correlations with other wavelength bands such as radio continuum and X--rays. 

In the case of NGC\,891 a `thick disk' has been discovered by radio continuum 
observations, that show a correlation with the eDIG in H$\alpha$ (Dahlem et 
al. \cite{DaDe}). Also some X--ray observations show a correlation, wherby 
here the hot ionized gas (HIM) is traced. The X--ray halo emission is a 
result of the interaction between the gas flows of supernovae explosions 
and/or star winds that are expelled from the starforming regions in the disk, 
which interact with the surrounding medium in the halo. These observations 
have aimed at the spatial extend which can be compared with the eDIG in 
optical observations. (e.g.,\, Bregman \& Pildis \cite{BrPi}; Fabbiano et al. 
\cite{FaHe}; Dahlem et al. \cite{Da98}). Moreover the dust features seen in 
several edge--on galaxies at high galactic latitudes with typically $z \sim$ 
300--1000 pc (e.g.,\,Howk \& Savage \cite{HoSa}, \cite{HoSa99}; 
Rossa \& Dettmar, in prep.) can also be compared with the DIG distribution. 


\section{Observations and Data Reduction}

\subsection{H$\alpha$ imaging}

The basis of our H$\alpha$ survey consist of 9 galaxies, for which data 
have been obtained in two observing runs with two different instruments. 
Optical H$\alpha$ narrow--band images of 6 edge--on spiral galaxies have 
been obtained with the ESO Faint Object Spectrograph Camera 2 (EFOSC2) in  
imaging mode, attached to the ESO/MPI 2.2m telescope at La Silla, Chile on 
Feb. 20--21 1993. The used CCD was the ESO CCD\#19 TH--chip with a pixel array 
of 1024$\times$1024 pixel. The pixel scale is $0\farcs34\,\rm{pix^{-1}}$. The 
narrowband images were taken through the ESO H$\alpha$ filters No. 694, 697, 
and 439. The equivalent widths of the filters are 32.4, 32.6, 44.3{\AA} 
respectively. The total integration times of the H$\alpha$ images were 
3600\,sec. on average, splitted into two images. The journal of the 
observations is given in Table~\ref{T1}. Additional R--band images have been 
obtained in order to perform a continuum subtraction. The integration times 
were 10--15 minutes for each galaxy in the R--band.  

In addition a small sample consisting of 4 edge--on spirals have been 
observed with the ESO Multi Mode Instrument (EMMI) at the NTT in imaging 
mode. The observations have been carried out during May 7--8 1991. In this 
observing campaign the ESO CCD--chip \#24 has been used. The FA--2048--L 
chip has a pixel array of 2048$\times$2048 pixel. The achieved pixel scale is 
$0\farcs27\,\rm{pix^{-1}}$. The H$\alpha$ images were all taken through the 
ESO filter No. 595. For the H$\alpha$ images the integration times were 
1800\,sec and 3600\,sec. The R--band integration times were 600\,sec each 
(see Table~\ref{T1}). Since the edge--on spiral IC\,2531 has been observed 
in both observing runs, 9 galaxies in total have been investigated.
  
\begin{table*}
\caption[]{Journal of observations}
\label{T1}
\begin{flushleft}
\begin{tabular}{llllllll}
\noalign{\smallskip}
\hline
Galaxy & Date & Instrument & H$\alpha$ filter $\lambda_c$ [{\AA}] & FWHM 
[{\AA}] & H$\alpha$ exposures & R-band exposures & Seeing \\ 
\hline
\noalign{\smallskip}
\object{NGC\,1963} & 20/02/1993 & EFOSC\,II & 6571.86 & 61.22 & 2 $\times$ 
1800\,sec & 900\,sec & $1\farcs2$ \\
\object{IC\,2531} & 08/05/1991 & EMMI & 6607.14 & 71.32 & 1 $\times$ 
1800\,sec & 600\,sec & $1\farcs2$ \\ 
IC\,2531 & 21/02/1993 & EFOSC\,II & 6651.69 & 61.30 & 2 $\times$ 1800\,sec & 
900\,sec & $1\farcs4$ \\
\object{NGC\,3044} & 20/02/1993 & EFOSC\,II & 6571.86 & 61.22 & 2 $\times$ 
1800\,sec & 900\,sec & $1\farcs1$ \\
\object{NGC\,4302} & 21/02/1993 & EFOSC\,II & 6571.86 & 61.22 & 2 $\times$ 
1800\,sec & 900\,sec & $1\farcs0$ \\
\object{NGC\,4402} & 20/02/1993 & EFOSC\,II & 6571.86 & 61.22 & 2 $\times$ 
1800\,sec & 900\,sec & $1\farcs1$ \\
\object{NGC\,4634} & 20/02/1993 & EFOSC\,II & 6571.86 & 61.22 & 2 $\times$ 
1800\,sec & 900\,sec & $1\farcs1$ \\
\object{NGC\,5170} & 07/05/1991 & EMMI & 6607.14 & 71.32 & 1 $\times$ 
1800\,sec & 600\,sec & $0\farcs8$ \\
\object{IC\,4351} & 08/05/1991 & EMMI & 6607.14 & 71.32 & 2 $\times$ 
1800\,sec & 600\,sec & $1\farcs0$ \\
\object{UGC\,10288} & 08/05/1991 & EMMI & 6607.14 & 71.32 & 2 $\times$ 
1800\,sec & 600\,sec & $0\farcs9$ \\
\noalign{\smallskip}
\hline
\end{tabular}
\end{flushleft}
\end{table*}

\subsection{Data reduction}

The data reduction was performed in the usual manner using the IRAF\footnote{
IRAF is distributed by the National Optical Astronomy Observatories, which is 
operated by the Association of Universities for Research in Astronomy, Inc. 
(AURA) under cooperative agreement with the National Science Foundation.} 
packages, including bias level correction, flat--fielding in order to remove 
the sensitivity variations. The images have been background corrected. This 
was done by measuring the intensity of various regions in the CCD image field 
that were neither contaminated by galaxy emission nor by bright stars, that 
contribute to a certain level to the background intensity. The field size of 
the boxes chosen for the background subtraction was typically $\sim$50\,pix$
\times$30\,pix. Usually three different fields were measured, to get a median 
level for the background, which was subtracted for each galaxy frame. 

In order to study the H$\alpha$ emission in the galaxies, the continuum 
emission in the filter passband has to be corrected. For this purpose the 
R--band images had to be scaled and subtracted from the H$\alpha$ images. 
This was done in the following way. First the R--band and H$\alpha$ line 
images had to be aligned. For that purpose the pixel coordinates of three to 
four stars in each frame have been measured and the H$\alpha$ frames have 
finally been shifted accordingly. Then the countrates of a region in the 
galaxy which was considered free from H$\alpha$ emission have been determined 
in both the R and H$\alpha$ images. The ratio of the two determined values is 
the scaling factor. The R-band image was divided by that value and then 
subtracted from the H$\alpha$ image. By that procedure a continuum free 
H$\alpha$+[\ion{N}{ii}] image is created. Finally the two H$\alpha$ images 
have been combined to gain a better S/N ratio. Some cosmics have been removed 
manually. The detection of cosmics was straight forward since we had two 
H$\alpha$ images for each galaxy, which can be easily compared. 

Finally a flux calibration has been performed. The measured R--band 
photometry data of our edge--on galaxies have been taken from 
NED\footnote{NASA Extragalactic Database}. These R--band magnitudes represent 
measured magnitudes in a certain aperture. Therefore we determined
the flux (countrates) in our R--band image for the same aperture size as were 
the photometry data, applying the flux of the standard Vega as a calibrator. 
We then divided the calculated flux by the value of the measured countrates in 
our R--band filter, correcting for the integration time and the scaling 
factor of our passband exposures. Finally we multiplied this factor with 
the measured countrate in our H$\alpha$ exposure and thus get the relation 
between the countrates and the flux. The uncertainties with this method are 
of the order of 20\%. 


\section{Analysis}

With the transformed flux units it is also possible to convert the flux 
values to another commonly used unit, namely the emission measure (EM) which 
is defined by 

\begin{equation}
EM = \int^r_0 n_{\rm{e}}^2\,dl
\end{equation}

\noindent The emission measure can be calculated according to 

\begin{equation}
EM = 2.75 \times T_4^{0.9}\,I(\rm{H}\alpha)\,\rm{cm^{-6}\,pc}
\end{equation}

\noindent with $I$(H$\alpha$) the intensity of the H$\alpha$ emission, in 
Rayleigh (R), and with $T$ the gas temperature in units of $10^4$\,K. 
(Reynolds \cite{Re90}), assuming case B photionisation (Osterbrock 
\cite{Os89}). Usually a conversion is derived at H$\alpha$ where $\rm{1\,cm^
{-6}\,pc = 2.06 \times 10^{-18} erg\,s^{-1}\,cm^{-2}\,arcsec^{-2}}$.
 
Although the continuum has been subtracted from the images, there is still 
some contamination of the H$\alpha$ line from the nearby [\ion{N}{ii}] 
doublet, which also contributes to the line emission. This is due to the given 
filter passband. With knowledge of line ratios, derived from spectroscopical 
investigations, the mean ratio of H$\alpha$ to [\ion{N}{ii}] can in principle 
be determined, which in turn can be used to correct for the [\ion{N}{ii}] 
emission. From the measured H$\alpha$ line flux the total H$\alpha$ luminosity 
can be computed by 

\begin{equation}
L_{\rm{H}\alpha} = 4\,\pi\,D^2\,F_{\rm{H}\alpha}
\end{equation}

\noindent where $D$ is the distance to the galaxy. It is now possible to 
derive the star formation rate ($SFR$) using the calibration of Madau et al. 
(\cite{Mad98}) to a Salpeter initial mass function (IMF) with mass limits 
0.1 and 100 $\rm{M}_{\sun}$ (Salpeter \cite{Sal55}) which after Kennicutt 
(\cite{Ke98b}) yields 

\begin{equation}
SFR\,[{\rm M_{\sun}\,yr^{-1}}] = 7.9 \times 10^{-42}\,L_{\rm{H}\alpha}\,
[\rm{erg\,s^{-1}}]. 
\end{equation}

\noindent The results can be compared with $SFR$s derived by far--infrared 
(FIR) fluxes from e.g., measurements with the IRAS satellite. The $SFR$, as 
derived from the FIR flux, are expected to be higher unless a correction 
of the H$\alpha$ fluxes is performed for internal dust absorption (since 
many edge--on galaxies bear a more or less prominent dust lane). The $SFR$ 
from the FIR luminosity can be calculated in a similar manner according to 
Kennicutt (\cite{Ke98b}), and references therein, taking into account the 
timescales for bursts of SF, which yields

\begin{equation}
SFR\,[{\rm M_{\sun}\,yr^{-1}}] = 4.5 \times 10^{-44}\,L_{\rm{FIR}}\,
[\rm{erg\,s^{-1}}] 
\end{equation}

\noindent which is actually valid for starburst galaxies. The sensitivity 
of our observations is $7.2 \times 10^{-18}\,\rm{erg\,s^{-1}\,cm^{-2}\,arcsec^{
-2}}$ on average, which corresponds to an emission measure (EM) of $3.5\,
\rm{cm^{-6}\,pc}$.


\section{Results}

\subsection{The sample}

Basic parameters for our sample of 9 edge--on galaxies are given in 
Table~\ref{T2}. Here the coordinates for the epoch J2000 along with 
morphological type, distances, heliocentric corrected radial velocities, 
sizes, inclinations, and R--band magnitudes are listed. The selection criteria 
for most of the objects of our sample were the following. Initially a list 
has been created from the Uppsala General Catalogue of Galaxies (UGC) 
(Nilson \cite{Ni73}) that has been used as a sample of edge--on galaxies for 
radio continuum observations (Hummel et al. \cite{HuBe}). The inclination 
criterium was $i \geq 75\degr$ in addition to the size criterium. Additional 
objects fulfilling the size and inclination criteria were observed to make 
optimal use of the Sidereal Time coverage during observations. 

A further selection criterium was to study nearby galaxies. All studied 
objects have distances of $D\leq$40\,Mpc, where the spatial resolution is 
sufficiently high to study morphological features (e.g.,\,plumes) that are 
related to a gas outflow from the disk into the halo. We are assuming a 
Hubble parameter of $\rm{H_0} = 75\,\rm{km\,s^{-1}\,Mpc^{-1}}$, that we will 
adopt throughout this paper. For the most distant galaxy in our sample 
(IC\,4351) $1\arcsec$ corresponds to 172\,pc, and for the nearest galaxy 
$1\arcsec = 83\,\rm{pc}$.
  
\begin{table*}
\setcounter{table}{1}
\caption[]{Basic galaxy parameters}
\label{T2}
\begin{flushleft}
\begin{minipage}{20cm}\small
\begin{tabular}{lllllllll}
\noalign{\smallskip}
\hline
Galaxy\footnote{All data have been taken or were calculated from the RC3 (de 
Vaucouleurs et al. \cite{Vau91}), except where indicated} & R.A. (J2000) & 
Dec. (J2000) & Type & $D$\,[Mpc] & $v_{\rm HI}\,[\rm{km\,s^{-1}}]$ & $a 
\times b$ & $i$ & $m_{\rm R}$\footnote{taken from NED, and from Schroeder \& 
Visvanathan \cite{ScVi}} \\
\hline
\noalign{\smallskip}
NGC\,1963 & $\rm{05^h33^m12\fs8}$ & $-36\degr23\arcmin59\arcsec$ & Sc & 
17.7 & 1324 & $3\farcm8\times0\farcm8$ & $84\fdg5$ & 12.11\\
NGC\,3044 & $\rm{09^h53^m39\fs8}$ & $+01\degr34\arcmin46\arcsec$ & SBb & 
17.2 & 1292 & $5\farcm7\times0\farcm6$ & $84\fdg0$ & 11.65\\
IC\,2531 & $\rm{09^h59^m55\fs7}$ & $-29\degr36\arcmin55\arcsec$ & Sc & 
33.0 & 2474 & $7\farcm5\times0\farcm9$ & $90\fdg0$ & 11.41\\
NGC\,4302 & $\rm{12^h21^m42\fs4}$ & $+14\degr36\arcmin05\arcsec$ & Sc & 
18.8\footnote{taken from Teerikorpi et al. \cite{TeBo}} & 1108$^{
\footnotesize{c}}$ & $5\farcm5\times1\farcm0$ & $88\fdg0$ & 12.11\\
NGC\,4402 & $\rm{12^h26^m07\fs9}$ & $+13\degr06\arcmin46\arcsec$ & Sb & 
22.0$^{\footnotesize{c}}$ & 237 & $3\farcm9\times1\farcm1$ & 
$83\fdg1$ & 12.09\\
NGC\,4634 & $\rm{12^h42^m40\fs4}$ & $+14\degr17\arcmin47\arcsec$ & Sc & 
19.1$^{\footnotesize{c}}$ & 1118$^{\footnotesize{c}}$ & 
$2\farcm6\times0\farcm7$ & $83\fdg0$ & 12.42\\
NGC\,5170 & $\rm{13^h29^m49\fs0}$ & $-17\degr57\arcmin59\arcsec$ & Sc & 
20.0 & 1503 & $8\farcm3\times1\farcm0$ & $83\fdg1$ & 10.77\\
IC\,4351 & $\rm{13^h57^m54\fs1}$ & $-29\degr18\arcmin54\arcsec$ & Sb & 
35.5 & 2662 & $5\farcm8\times1\farcm1$ & $78\fdg9$ & 11.18\\
UGC\,10288 & $\rm{16^h14^m25\fs1}$ & $-00\degr12\arcmin27\arcsec$ & Sc & 
27.3 & 2045 & $4\farcm7\times0\farcm5$ & $83\fdg4$ & 12.20\\
\noalign{\smallskip}
\hline
\end{tabular}
\end{minipage}
\end{flushleft}
\end{table*}

All 9 galaxies of our sample are late--type galaxies (Sb--Sc), and DIG 
has been detected in galaxies of this type before, as already mentioned in 
Sect.\,1. Five of them show an eDIG layer with extraplanar distances of $z 
\leq$ 2\,kpc above the galactic midplane, whereby 1 galaxy shows only 
plumes or filaments reaching into the halo. In Table~\ref{T4} we give a 
summary of the observed DIG features, that will be reviewed in detail for 
each galaxy below.

\subsection{Individual results for each galaxy}

In this section we present the results for each of our selected galaxies 
separately and discuss them in Sect.\,5. The field sizes of the figures 
have been chosen conveniently to show the larger galaxies entirely, except 
for IC\,2531 which did not fit completely in the field of view. The smaller 
ones are shown as enlargements to give more details. Spatial profiles of the 
DIG emission are presented in Fig.~\ref{F9}.

\vspace{0.2cm}
 
\begin{center} {\em NGC\,1963} \end{center}
This galaxy has been classified as type Sc (Lauberts \cite{La82}). NGC\,1963 
is poorly studied. It appears in the extended 12 micron galaxy sample (Rush 
et al. \cite{RuMa}) with measured fluxes of $F_{\rm{60\mu m}}$ = 2.99\,Jy and 
$F_{\rm{100\mu m}}$ = 7.38\,Jy. An elliptical galaxy is visible $\sim 
1\farcm5$ to the NNE. This is a member of the galaxy cluster \object{Abell 
S0535}, which is at a distance of z = 0.0473 (Quintana \& Ram\'{\i}rez 
\cite{QuRa}) and hence not associated with NGC\,1963. It is important to 
check the surrounding field of a galaxy for companion galaxies, as those 
galaxies -- if closeby to the parent galaxy (e.g.,\, same redshift) -- may 
also trigger gas outflows. Due to the scaling procedure this elliptical 
galaxy is not visible in the H$\alpha$ image (it looks like an outmasked 
image) as objects with no or almost non--detectable H$\alpha$ emission 
appear. (The two white vertical lines are dead--pixel columns).  

\vspace{0.2cm}
\begin{figure}[h]
\psfig{figure=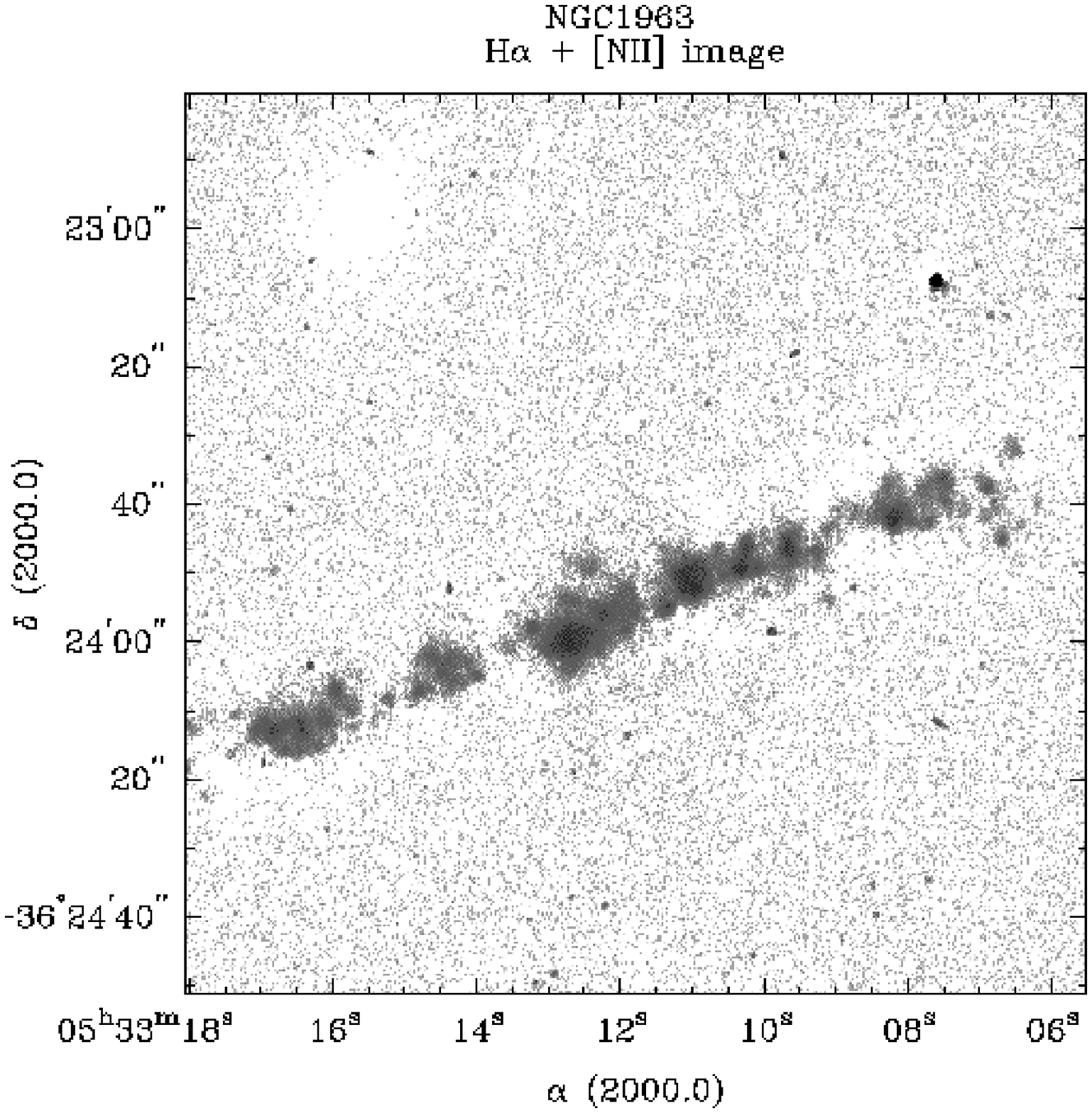,width=8.8cm,clip=t}
\caption[]{H$\alpha$+[\ion{N}{ii}] image of NGC\,1963. The scale is $1\arcsec 
= 86\,\rm{pc}$.}
\label{F1}
\end{figure}
\vspace{0.2cm}

A weak eDIG layer is seen (0.5--0.6\,kpc on average) as well as several 
extraplanar patches on either side of the galaxy disk. The most prominent 
feature is just north of the center ($R$=0.2\,kpc). It is located at 
$z \approx 0.7\,\rm{kpc}$ above the galactic plane. The extended emission is 
seen around several bright \ion{H}{ii} regions, that are not continuously 
distributed along the galaxy plane. They are more or less clustered in small 
groups. The H$\alpha$ flux of the galaxy is estimated to be $\rm{1.64 \times 
10^{-13}\,erg\,s^{-1}\,cm^{-2}}$ and from the derived H$\alpha$ luminosity the 
global star formation rate ($SFR$) has been determined which is $SFR = 
0.05\,\rm{M_{\sun}\,yr^{-1}}$.

The stars in the H$\alpha$ images usually appear as residuals due to 
the slightly different point spread function in the various filters. The 
resulting H$\alpha$ image is shown in Fig.~\ref{F1}.

\vspace{0.5cm}

\begin{center} {\em NGC\,3044} \end{center} 
There has been a slight controversy, whether this galaxy is a galaxy with 
enhanced star formation (SF), hence with higher SF than in `normal' 
(quiescent) galaxies, or whether it is a starburst galaxy. In a recent survey 
of DIG emission in edge--on starburst galaxies (Lehnert \& Heckman 
\cite{LeHe}) NGC\,3044 has also been included, while other researchers 
classify it as a non--starburst galaxy (Hummel \& van der Hulst \cite{HuHu}; 
Dahlem et al. \cite{DaLi}). However, the galaxy has been listed in the sample 
of IRAS bright galaxies (Soifer et al. \cite{So87}), so there seems clear 
evidence for enhanced SF but it is presently not clear if NGC\,3044 hosts a 
starburst nucleus. In a spectroscopic study where the DIG was investigated at 
two different slit positions perpendicular to the galaxy disk, the positions 
of the detected DIG in the diagnostic diagrams fall in between the areas 
occupied by normal \ion{H}{ii} regions and starburst (T\"ullmann \& Dettmar 
\cite{TuDe}). Therefore no clear answer of this debate can be given yet. 
Even the term `starburst' is sometimes not clearly defined, as various 
researchers use different definitions. We will come back to this point in 
Sect.\,5.

The galaxy type is listed as SBc (Tully \cite{Tu88}). Even classifications 
as a non--barred galaxy (Sc) are listed (Nilson \cite{Ni73}). There are 
indications that in NGC\,3044 a bar is present, and \ion{H}{i} kinematics 
by Lee \& Irwin (\cite{LeIr}) has indeed discerned a bar. It might be worth 
to notice that in the eighties a supernova (\object{SN1983E}) has been 
detected in this galaxy (Barbon et al. \cite{BaCa}).

\begin{figure}[h]
\psfig{figure=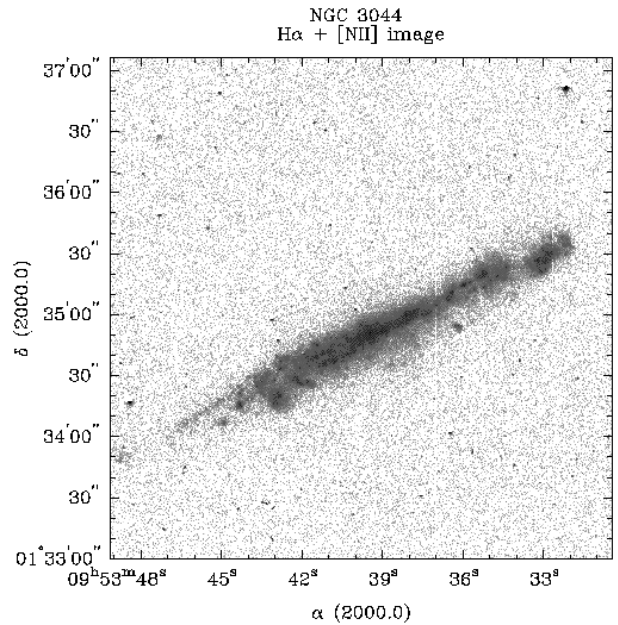,width=8.8cm,clip=t}
\caption[]{H$\alpha$+[\ion{N}{ii}] image of NGC\,3044. $1\arcsec$ corresponds 
to 83\,pc.}
\label{F2}
\end{figure}

In Fig.~\ref{F2} we present the H$\alpha$ image. The morphology of the DIG 
shows various features. An eDIG layer can be detected at extraplanar 
distances up to $z=0.8-1\,\rm{kpc}$. Several single plumes can also be 
discerned. South of the galactic plane an extended structure is visible, 
which has a loop--like appearance. This loop extends out to $\sim$ 1.8\,kpc 
with a radius of about 1\,kpc, and resembles the galactic supershells. The 
disk appears slightly warped, which is also apparent in the R--band image. 
The H$\alpha$ flux of the galaxy has been estimated to be $\rm{2.50 \times 
10^{-12}\,erg\,s^{-1}\,cm^{-2}}$ and from the computed H$\alpha$ luminosity 
the (global) star formation rate (SFR) has been derived, which is $\rm{SFR = 
0.71\,M_{\sun}\,yr^{-1}}$. 

\vspace{0.2cm}
\vspace{0.2cm}

\vspace{1cm}

\begin{figure}[h]
\psfig{figure=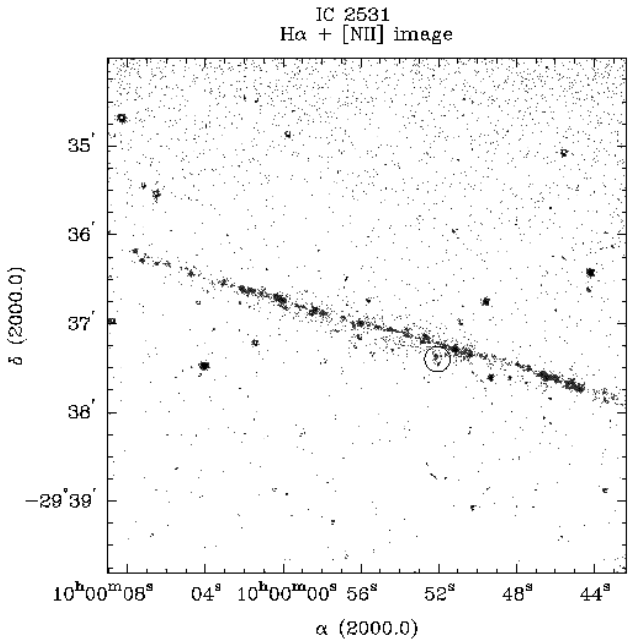,width=8.8cm,clip=t}
\caption[]{H$\alpha$+[\ion{N}{ii}] image of IC\,2531. The image measures 
$5\farcm8 \times 5\farcm8$, which is the field of view of EFOSC\,II in the 
used configuration. Therefore the galaxy disk is not covered entirely. 
$1\arcsec = 160\,\rm{pc}$.}
\label{F3}
\end{figure}

\begin{center} {\em IC\,2531} \end{center}
This southern edge--on spiral is slightly larger than the EFOSC2 field of view 
which is $5\farcm8 \times 5\farcm8$. IC\,2531 is seen perfectly edge--on. In 
our H$\alpha$ image almost no extraplanar diffuse emission has been detected. 
One filament (the chimney--like feature) is clearly seen, emerging from the 
disk radius at $R$=6\,kpc south of the plane into the halo (see Fig.~\ref{F3}
). This feature is marked in Fig.~\ref{F3} with a circle. It reaches a 
height of $z$=2\,kpc above the galactic plane. The H$\alpha$ image 
looks pretty much like a string of pearls. Several disk \ion{H}{ii} regions 
can be identified, but only the largest are surrounded by DIG, which is 
probably not extraplanar. Only at $R$=8.4\,kpc from the center a larger disk 
\ion{H}{ii} region seems to be embedded in a fainter DIG layer reaching an 
extraplanar height of $z$=1.1\,kpc.

\vspace{0.2cm}
\vspace{0.2cm}

\vspace{-0.5cm}
\begin{figure}[h]
\psfig{figure=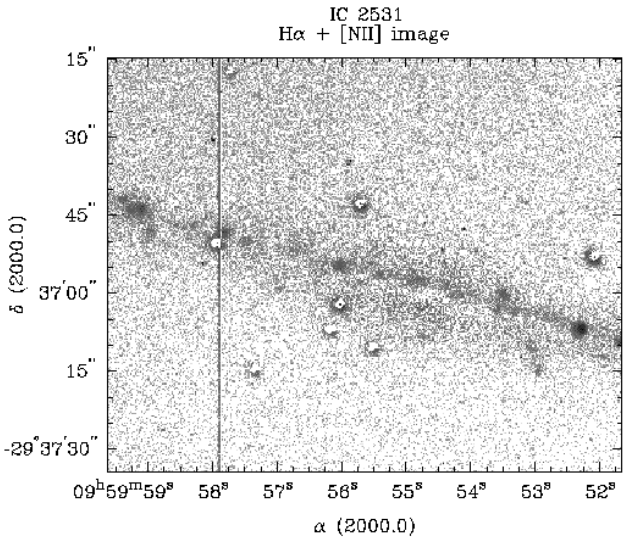,width=8.8cm,clip=t}
\caption[]{Central part of the H$\alpha$+[\ion{N}{ii}] image of IC\,2531, 
obtained with the NTT.}
\label{F4}
\end{figure}

There are no radio continuum observations available, and the only conspicous 
feature seen in the R-band images is a 'peanut-bulge'. IC\,2531 looks similar 
to the well studied edge--on spiral NGC\,891 in the optical. However, 
in H$\alpha$ they appear totally different. No Far--Infrared (FIR) emission 
has been detected with the IRAS satellite in IC\,2531. Therefore it can be 
concluded, that the $SFR$ in this edge--on spiral is completely different 
from that in NGC\,891. From a theoretical study in comparison with 
multi--color surface brightness profiles, observations to model the vertical 
structure of the disk it is found that the disk in IC\,2531 has a similar 
composition as the disk of our Milky Way (Just et al. \cite{JuFu}).  

In Fig.~\ref{F4} we show an enlargement of the central part of IC\,2531. This 
image has been obtained with the NTT. Here the filament south of the galaxy 
disk can be seen in a little more detail. The proof that this filament is 
real and not an artifact is given by its presence in both of our images, 
taken with different instruments. Furthermore we have always obtained two 
exposures for each galaxy with any given instrumental setup, where we can 
distinguish between faint emission and cosmics that sometimes can mimic faint 
emission. Since DIG is traced by H$\alpha$ emission, it is evident that there 
is low SF activity in the disk of IC\,2531. 

\vspace{0.5cm}

\begin{figure}[h]
\psfig{figure=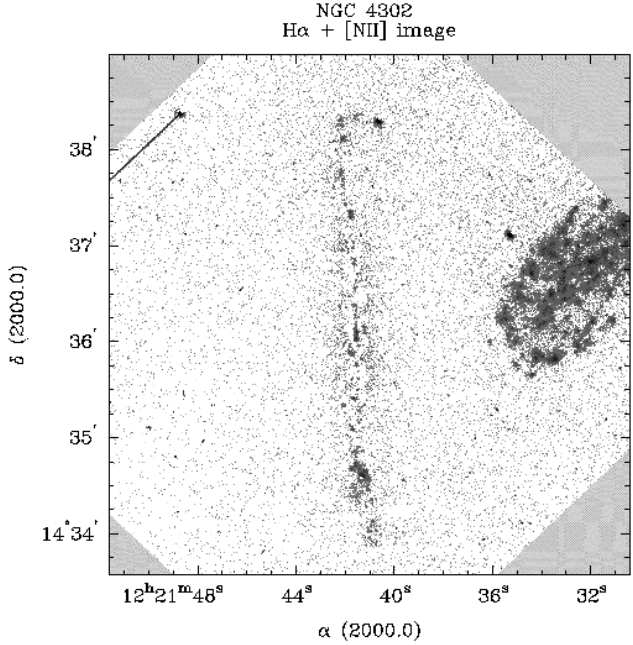,width=8.8cm,clip=t}
\caption[]{H$\alpha$+[\ion{N}{ii}] image of NGC\,4302. The face--on spiral on 
top right is NGC\,4298. $1\arcsec = 91\,\rm{pc}$.}
\label{F5}
\end{figure}

\begin{center} {\em NGC\,4302} \end{center}
This edge--on galaxy has already been studied before in the DIG context with 
H$\alpha$ imaging by two different groups (Pildis et al. \cite{Pi94}; 
Rand \cite{Ra96}). In our H$\alpha$ we see faint eDIG emission, when averaging 
the intensities perpendicular to the galaxy disk. This is consistent with the 
results from Rand (\cite{Ra96}), who also detected eDIG. The single plume, 
already detected by Pildis et al. (\cite{Pi94}), is also visible in our image. 
The halo emission in NGC\,4302 is fainter than in NGC\,3044. However, the 
halo emission is much brighter than the emission from the disk. This is 
due to the extended dust lane (which is very prominent in NGC\,4302) along 
the disk. The dustlane absorbs most of the disk emission. Most of the other 
galaxies in our sample are not as much influenced by thick dust lanes as it 
is the case in NGC\,4302. The dust lanes are best visible in broadband images. 
In our H$\alpha$ image NGC\,4302 (see Fig.~\ref{F5}) is shown with a companion 
galaxy, the face--on spiral \object{NGC\,4298}. Whether this galaxy is 
capable to trigger the star formation in NGC\,4302 is not known yet. 

\begin{figure}[h]
\psfig{figure=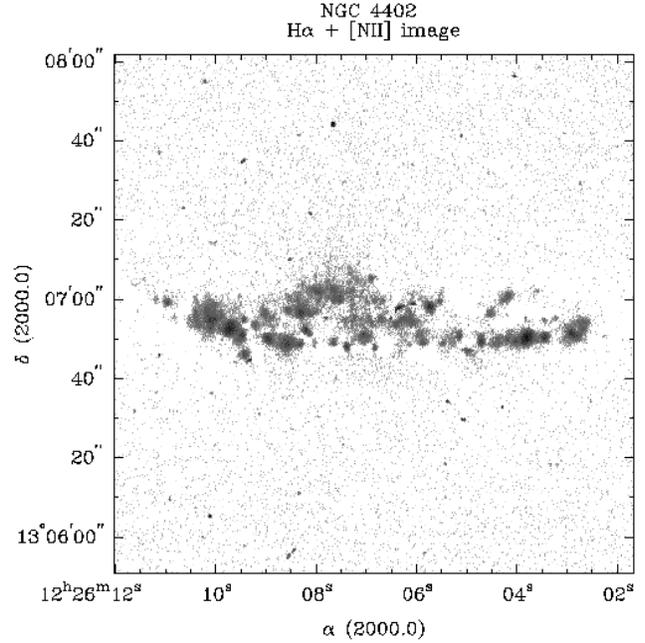,width=8.8cm,clip=t}
\caption[]{H$\alpha$+[\ion{N}{ii}] image of NGC\,4402. The scale is $1\arcsec 
= 107\,\rm{pc}$ }
\label{F6}
\end{figure}

\begin{center} {\em NGC\,4402} \end{center}
In this edge--on spiral an eDIG is detected showing extraplanar distances of 
$\sim$ 2.2\,kpc. The DIG emission is restricted to the eastern part of the 
galaxy. Small filaments are emanating the galaxy north of the galactic plane. 
Parts of the spiral structure can be discerned in the H$\alpha$ image due to 
the deviation from the exact edge--on sightline of $\Delta i \sim 7^\circ$. 
NGC\,4402 is a member of the Virgo cluster (cf. Binggeli et al. \cite{BiSa}). 
Dozens of relatively bright \ion{H}{ii} regions are embedded in DIG. This DIG 
is supposed to originate from emission that is leaking through the \ion{H}{ii} 
regions heated by hot O and B stars. The H$\alpha$ image is shown in 
Fig.~\ref{F6}. The visual appearance of NGC\,4402 is similar to the recently 
studied Virgo cluster member galaxy NGC\,4522 (Kenney \& Koopmann \cite{
KeKo}). They claim that they have detected eDIG up to 3\,kpc above the 
galactic plane. However, the inclination of that galaxy is far away from 
being edge--on, so the observed emission might be emission from the disk.

\vspace{0.2cm}
\vspace{0.2cm}

The H$\alpha$ flux of NGC\,4402 has been estimated and has a value of 
$\rm{1.47 \times 10^{-13}\,erg\,s^{-1}\,cm^{-2}}$ and from the derived 
H$\alpha$ luminosity the star formation rate ($SFR$) has been determined 
which is $SFR = \rm{0.07\,M_{\sun}\,yr^{-1}}$.

\vspace{0.2cm}
\vspace{0.2cm}
  
\begin{center} {\em NGC\,4634} \end{center}
This edge--on galaxy is also a member of the Virgo cluster (cf.\,Binggeli 
et al. \cite{BiSa}; Helou et al. \cite{HeHo}) and Teerikorpi et al. 
(\cite{TeBo}) give a distance of 19.1\,Mpc to this galaxy. Recently new 
R--band photometry became available for this galaxy and other members of the 
Virgo Cluster (Schroeder \& Visvanathan \cite{ScVi}). Oosterloo \& Shostak 
(\cite{OoSh}) list it as a binary pair with \object{NGC\,4633} in their 
\ion{H}{i} study. NGC\,4634 shows an interesting DIG morphology. A bright 
eDIG layer is detected, which reaches distances of $\sim$1.1\,kpc above the 
galactic plane. We show our H$\alpha$+[\ion{N}{ii}] image in Fig.~\ref{F7}. In 
addition to the eDIG layer, several filaments reach into the halo, similar to 
the ones discovered in the edge--on spiral \object{NGC\,5775} (Dettmar 
\cite{De93}). These plumes are more frequent than in NGC\,5775, although 
fainter in intensity on average. 

Furthermore several \ion{H}{ii} regions in the disk can be identified, and in 
the south--eastern part a possible dust region absorbs parts of the emission 
from the disk. Interestingly the visible \ion{H}{ii} regions in the disk seem 
not aligned in a plane. This might be an orientation effect due to the galaxy 
inclination of $83\degr$. However, the position of the \ion{H}{ii} regions 
along the plane scatter randomly, which might indicate that the disk is 
disturbed. Therefore an interaction of NGC\,4634 with neighbouring galaxies 
in the Virgo cluster seems likely.

\vspace{0.2cm}
\begin{figure}[]
\psfig{figure=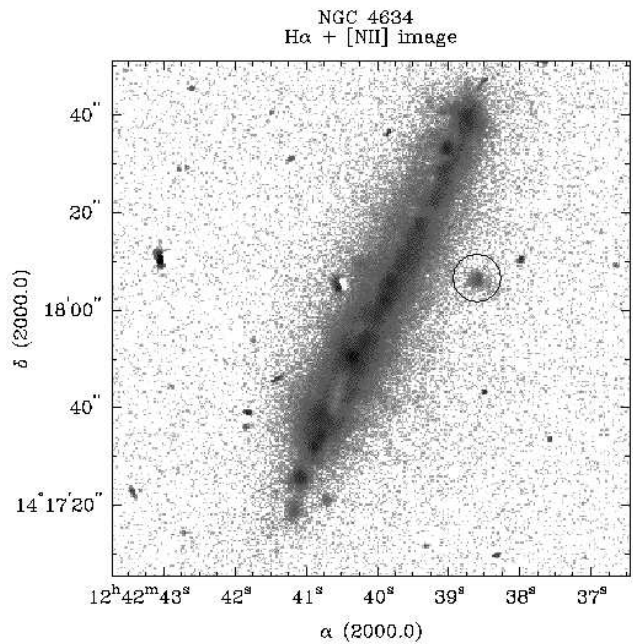,width=8.8cm,clip=t}
\caption[]{H$\alpha$+[\ion{N}{ii}] image of NGC\,4634. A bright eDIG layer 
is clearly visible. The extraplanar emission region (Patch\,1) is marked by 
a circle. The scale is $1\arcsec = 93\,\rm{pc}$.}
\label{F7}
\end{figure}
\vspace{0.2cm}

In the vicinity of NGC\,4634 another Virgo spiral is located, namely 
NGC\,4633. Both galaxies have similar radial velocities ($\rm{\Delta v 
\approx 112\,km\,s^{-1}}$), therefore a direct interaction seems very likely. 
The presence of a bar is reported in the literature, which could also be a 
source of disturbance.

About 10 bright \ion{H}{ii} regions can be identified in the disk which are 
embedded in DIG. Some of the filaments reaching into the halo, which are seen 
in our H$\alpha$ image, can be traced back to the disk. This would be 
consistent with theoretical models ('chimneys'), with the chimneys as the 
interface between disk and halo. At least one \ion{H}{ii} region protudes 
from the disk. This is the second bright emission region in the very northern 
part of the disk, which is slightly offset from the disk. 

The most outstanding feature is seen NW of the galaxy disk in the halo. This 
feature, which we refer to as Patch\,1, is a small isolated emission patch, 
that is clearly visible in our H$\alpha$ images, located $\sim$1.4\,kpc above 
the galactic plane. Whether this prominent feature is related to the eDIG is 
not clear yet. Additional longslit spectroscopy is necessary to reveal the 
true nature of this object, whether it is truly related to the eDIG or rather 
a (projected) dwarf galaxy. However, if this morphological feature, which 
seems to show no direct connection (at the faint level) to the disk, will 
indeed be confirmed spectroscopically (e.g.,\,same redshift) as part of the 
eDIG in NGC\,4634, this would give rise to a new phenomenon visible in halos 
of edge--on galaxies which might be coined as {\em star formation in galactic 
halos}. No object was found in a search of the NED and SIMBAD databases at 
this position. 

In Fig.~\ref{F9} we present cuts perpendicular to the major axis of our 
studied edge--on galaxies. Typically 30 pixel scans have been averaged along 
the minor axis. In these spatial profiles the H$\alpha$ flux in $\rm{erg\,
s^{-1}\,cm^{-2}}$ is plotted as a function of the spatial coordinate ($z$). 
For NGC\,4634 the highest peak corresponds to the disk and halo region and 
the secondary peak to the right is the emission patch Patch\,1, located about 
1.4\,kpc above the galactic plane. The H$\alpha$ flux of NGC\,4634 has been 
estimated, without correcting for internal absorption, to be $\rm{1.47 \times 
10^{-12}\,erg\,s^{-1}\,cm^{-2}}$ and from the computed H$\alpha$ luminosity 
the SFR has been derived ($SFR = 0.51\,\rm{M_{\sun}\,yr^{-1}}$). 

A radio continuum flux (flux density) at $\lambda$ = 2.8\,cm of 6 $\pm$1\,mJy 
and at $\lambda$ = 6.3\,cm of 20 $\pm$2\,mJy has been reported (Niklas et al. 
\cite{NiKl}) but no maps are shown, which would allow a comparison between 
radio continuum and H$\alpha$. No X--ray observations have been performed up 
to now to study the hot ionized gas.  

\vspace{0.2cm}
\begin{figure*}[]
\setcounter{figure}{8}
\psfig{figure=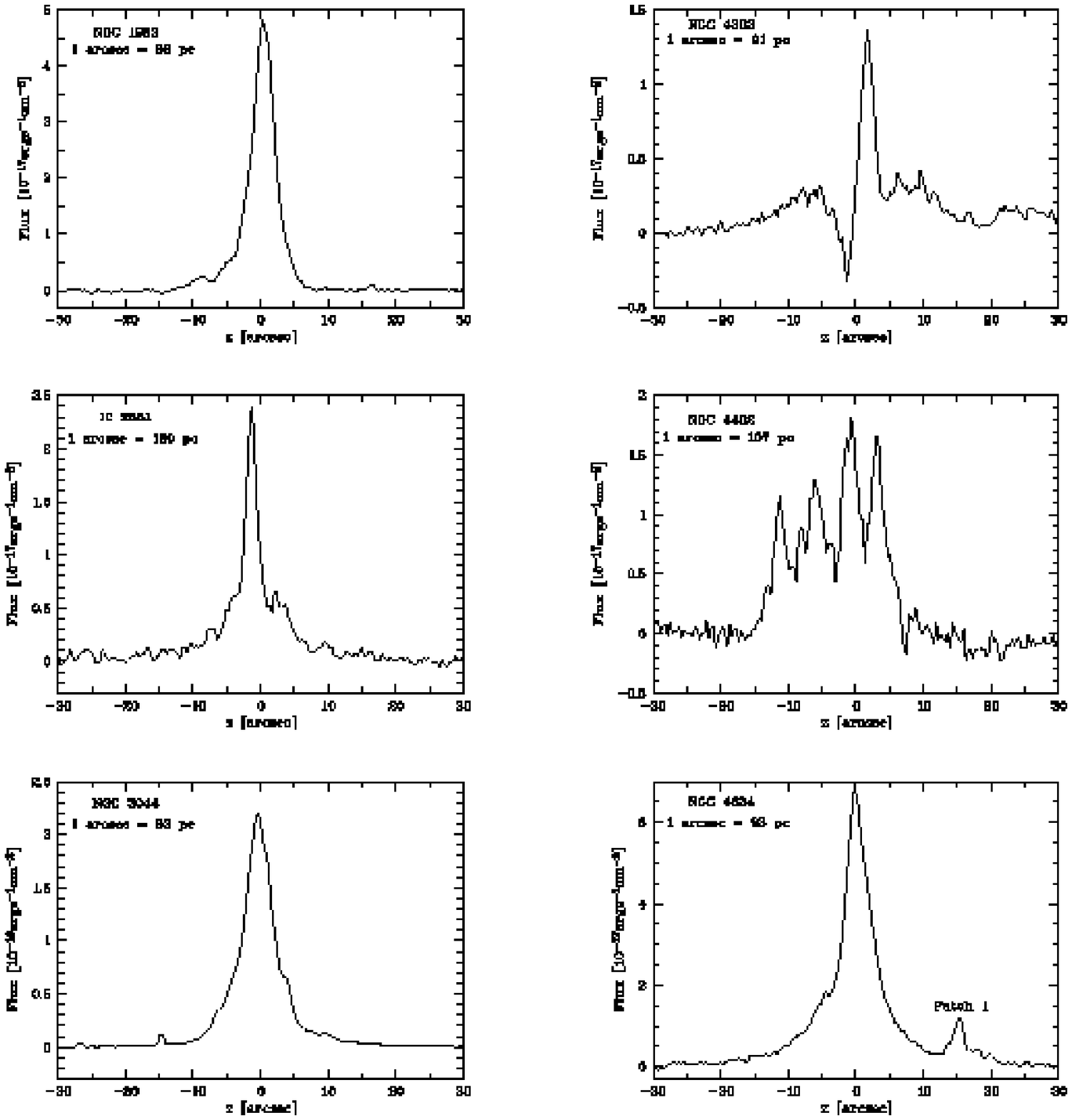,width=16.8cm,clip=t}
\caption[]{Spatial profiles of the H$\alpha$ emission in 6 of our studied 
edge--on galaxies (cuts perpendicular to the major axis). In each case several 
pixel rows (NGC\,1963: 34, IC\,2531: 22, NGC\,3044: 30, NGC\,4302: 78, 
NGC\,4402: 31, and NGC\,4634: 21) have been averaged around representative 
regions. The flux (in $\rm{erg\,s^{-1}\,cm^{-2}}$) is plotted as a function 
of the distance from the galactic plane ($z$) in arcseconds.}
\label{F9}
\end{figure*}
\vspace{0.2cm}

\vspace{0.5cm}

\begin{center} {\em NGC\,5170} \end{center}
This is another example of a poorly studied southern spiral galaxy. 
No eDIG is detected in this edge--on galaxy. Our image shows the bulge 
slightly oversubtracted. From the FIR flux one can expect not to have high 
$SFR$s, if one assumes a correlation between the FIR luminosity and the $SFR$. 
However, some galaxies seem to have high local $SFR$ but not on a global 
scale. Therefore for an intrinsically large galaxy the integrated FIR would be 
lower, and a correlation with the FIR flux seems not always apropriate as a 
tracer for high $SFR$ and hence for gas outflows from the disk into the halo. 
A better indicator for localized SF activity, which is often used, is the 
ratio of the FIR luminosity and the isophote diameter with $25^{\rm th}$\,
mag/$\sq\arcsec$ [$L/D_{25}^2$] (e.g.,\,Rand \cite{Ra96}). We will discuss 
this point in Sect.\,5 in detail. The H$\alpha$ image is shown in 
Fig.~\ref{F8}. 

\vspace{0.2cm}
\begin{figure}[h]
\setcounter{figure}{7}
\psfig{figure=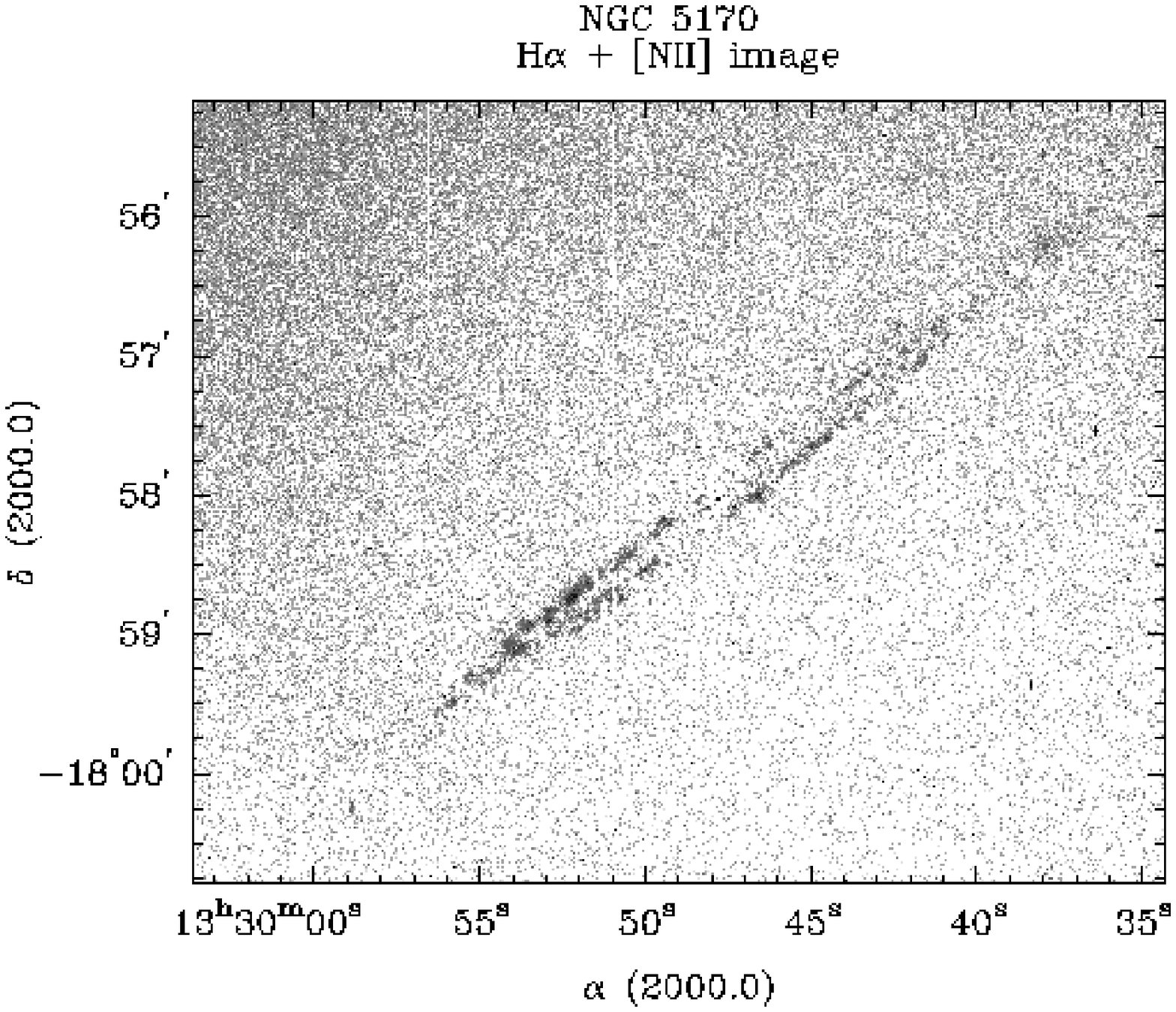,width=8.8cm,clip=t}
\caption[]{H$\alpha$+[\ion{N}{ii}] image of NGC\,5170. At the distance of 
NGC\,5170, $1\arcsec$ corresponds to 97\,pc.}
\label{F8}
\end{figure}
\vspace{0.2cm}

\begin{center} {\em IC\,4351} \end{center}
This southern edge--on spiral looks rather inconspicuous in H$\alpha$ (see 
Fig.~\ref{F10}). No eDIG is detected at the sensitivity level of our data. 
This is the only galaxy in our sample with an inclination slighly smaller than 
$80\degr$, where the spiral pattern becomes partly visible. In  starburst 
galaxies of similar inclination eDIG has been discovered. However, in 
IC\,4351 no morphological features can be discerned which might be related to 
eDIG. This is a counter--example of a Sb galaxy in the eDIG context. This 
reflects that not all late--type spiral galaxies bear disk--halo interaction 
(DHI) at a noticable level. 

\vspace{0.2cm}
\begin{figure}[]
\setcounter{figure}{9}
\psfig{figure=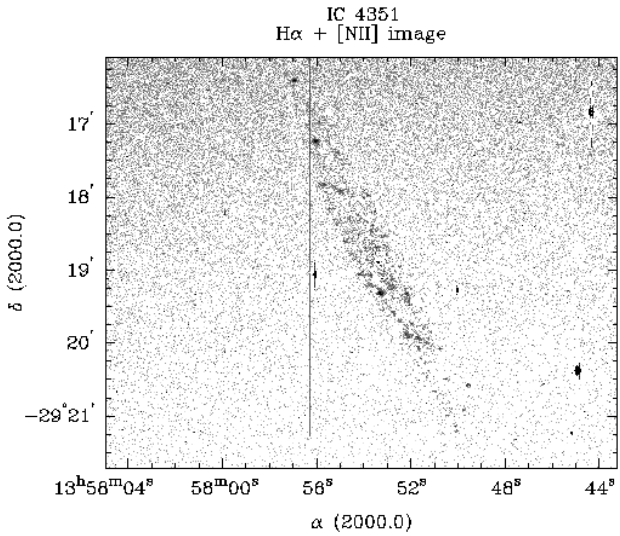,width=8.8cm,clip=t}
\caption[]{H$\alpha$+[\ion{N}{ii}] image of IC\,4351. At the distance of 
IC\,4351, $1\arcsec$ corresponds to 172\,pc.}
\label{F10}
\end{figure}
\vspace{0.2cm}

\begin{center} {\em UGC\,10288} \end{center}
Already studied by Rand (\cite{Ra96}), this galaxy has also been a target 
object of our survey. Unfortunately the H$\alpha$ image suffers from a bad 
S/N ratio and thus shows only little information. Therefore it is not 
reproduced here. Only the brightest \ion{H}{ii} regions in the disk are 
visible, but no further information can be retrieved from this image.

\subsection{Comparison of DIG morphology in our sample}

In Table~\ref{T4} we summarize the detections of the eDIG and list the 
morphological features for each galaxy. The morphology of the disk, as well 
as the halo, is presented in detail. We give the position in the ($R,z$) 
coordinates, with $R$ the coplanar distance from the nucleus, and $z$ the 
extraplanar distance. Furthermore we list references to radio continuum 
detections (e.g.,\,`thick disks') and correlations. The references of radio 
continuum observations are cited in Sect.\,4.2 for each galaxy. Moreover we 
list the FIR fluxes, as measured with the IRAS satellite, and give the 
computed star formation rates ($SFR$), that have been derived from our 
H$\alpha$ observations.

The determination of the SFRs in edge--on galaxies can generally (i.e. in the 
case of optical observations) only give lower limits of the true SFRs. This 
is basically due to internal absorption (extinction by dust), which can be 
quite prominent for edge--on galaxies with a strong dust lane, such as 
NGC\,891 (Dettmar \cite{De90}). The derived $SFR$s in our galaxy sample are 
up to 10 times lower than the $SFR$s calculated from the FIR flux, which 
would be expected, since we did not apply an extinction correction. Thus our 
derived $SFR$s from the H$\alpha$ flux are meant to be lower limits and for 
comparison between objects only.

The derived H$\alpha$ luminosities of our sample galaxies with clear 
eDIG detections (which range from $L_{\rm{H\alpha}}=6.2\times10^{39}\,
\rm{erg\,s^{-1}}$ to $L_{\rm{H\alpha}}=8.9\times10^{40}\,\rm{erg\,s^{-1}}$) 
are comparable to other spiral galaxies of the same Hubble type, which were 
studied for instance by Kennicutt \& Kent (\cite{KeKe83}). The derived 
H$\alpha$ flux of NGC\,3044 is also in agreement with the measurement by 
Lehnert \& Heckman (\cite{LeHe}).  

\begin{table*}
\setcounter{table}{3}
\caption[]{Summary of eDIG detections and comparision with radio continuum 
observations}
\label{T4}
\begin{flushleft}
\begin{minipage}{20cm}\small
\begin{tabular}{llllll}
\noalign{\smallskip}
\hline
Galaxy & DIG morphology\footnote{results from this investigation} & $R\,[''$], 
$z$\,[kpc]\footnote{coplanar and extraplanar distances of eDIG from the galaxy 
center} & $SFR$\,[$\rm{M_{\odot}\,yr^{-1}}$]\footnote{SFRs have been 
calculated according to Kennicutt (\cite{Ken98}) using our derived H$\alpha$ 
luminosities ($L_{\rm{H}\alpha}$)} & radio cont.\footnote{detections (yes), 
non-detections, or not yet observed (no), and correlations with H$\alpha$ 
(corr.) from various radio continuum\\ 
\hspace*{0.45cm} surveys (e.g.,\,Hummel et al. 1991)} & $\log 
L_{\rm FIR}$\footnote{FIR luminosities have been calculated from the FIR flux 
values given by Fullmer \& Lonsdale (1989). No FIR luminosity\\
\hspace*{0.45cm} for IC\,2531 could be calculated since no FIR flux 
measurements were available.}\\ 
\noalign{\smallskip}
\hline
NGC\,1963 & disk: bright \ion{H}{ii} regions & & 0.05 & no & 43.143 \\
 & halo: diffuse extended emission, plumes & 7W , 0.99 & & & \\
IC\,2531 & disk: several \ion{H}{ii} regions & & $\leq$0.13 & no &  \\
 & halo: one filament & 37W , $\sim2.0$ & & & \\
NGC\,3044 & disk: $\sim$ a dozen bright \ion{H}{ii} regions & & 0.71 
& yes/corr. & 43.606 \\
 & halo: diffuse, plumes, loop & N/S layer: $1.1-1.8$ & & & \\
NGC\,4302 & disk: absorption by dust, \ion{H}{ii} regions & & 0.13 & yes/corr. 
& 43.523 \\
 & halo: weak extraplanar layer + patches & $z\sim 0.8-0.9$ & & & \\
NGC\,4402 & disk: \ion{H}{ii} regions embedded in DIG & & 0.07 & yes & 
43.686 \\
 & halo: diffuse eDIG (localized) & 16E , 2.2 & & & \\
NGC\,4634 & disk: $\sim$ 10 bright \ion{H}{ii} regions & & 0.51 & yes & 
43.383 \\
 & disk-halo interface: bright \ion{H}{ii} region & 29W , 0.25 & & & \\
 & halo: diffuse, filaments & layer, $\sim$1.1 & & & \\
 & halo: Patch\,1 & 15W , 1.4 & & & \\
NGC\,5170 & disk: \ion{H}{ii} regions & & & yes & 42.898 \\
 & halo: no eDIG & -- -- -- --  & & & \\
IC\,4351 & disk: \ion{H}{ii} regions & & & no & 43.759 \\
 & halo: no eDIG & -- -- -- -- & & & \\
UGC\,10288 & disk: a few \ion{H}{ii} regions & & & yes & 43.108 \\
 & halo: no eDIG (due to insuff. S/N) & -- -- -- -- & & & \\
\noalign{\smallskip}
\hline
\end{tabular}
\end{minipage}
\end{flushleft}
\end{table*}

\subsection{Determination of the DIG properties $n_{\rm e}$ and $M$ }

In the previous sections the morphology of the DIG in our studied galaxies 
has been discussed. We can now derive some physical quantities that are 
directly related to the outflowing gas. We want to address the question on 
which percentage of the observed DIG is actually belonging to the halo, and 
which fraction is related to the disk (i.e. the diffuse gas component of the 
\ion{H}{ii} regions in the disk). We therefore determined the H$\alpha$ flux 
in our galaxies in various intervalls of extraplanar distances. First we 
determined the flux in the disk for which we assume a maximal extension of 
300\,pc from the galactic plane on either side of the galaxy disk. We also 
measured the total flux in the halo, for which we assume an extraplanar 
distance from 300 up to the distance of the most distant halo feature in each 
galaxy that is visible in our images. The comparison with the total flux 
yielded the respective fractions, which are summarized in Table~\ref{T3}.

\begin{table}[h]
\setcounter{table}{2}
\caption[]{Fractions of DIG distributions}
\label{T3}
\begin{flushleft}
\begin{minipage}{20cm}\small
\begin{tabular}{llll}
\noalign{\smallskip}
\hline
Galaxy & $f_{halo}(\rm{H}\alpha\,flux)$ & $f_{disk}(\rm{H}\alpha\,flux)$ & 
$\frac{f_{\rm halo}}{f_{\rm disk}}$
\\ \hline
NGC\,1963 & $12 \pm 3 \%$ & $88 \pm 22 \%$ & 0.136 \\
NGC\,3044 & $41 \pm 10 \%$ & $59 \pm 15 \%$ & 0.695 \\
NGC\,4302 & $59 \pm 15 \%$ & $41 \pm 10 \%$ & 1.439 \\
NGC\,4402 & $45 \pm 11 \%$ & $55 \pm 14 \%$ & 0.818 \\
NGC\,4634 & $36 \pm 9 \%$ & $64 \pm 17 \%$ & 0.563 \\
\noalign{\smallskip}
\hline
\end{tabular}
\end{minipage}
\end{flushleft}
\end{table}

We have calculated some physical quantities which make it possible to 
derive the mass of the DIG. First we determined the electron density 
of the diffuse halo gas by combining the equations (1) and (2). This yields 
with an assumed filling factor of $f=0.2$ (Dettmar \cite{De92})
\begin{equation}
n_e = \sqrt{2.75 \frac{T_4^{0.9}}{f\,D}\,\frac{F_{\rm H\alpha}}{\Omega}}
\end{equation} 

With the knowledge of $n_{\rm e}$ it is now possible to give estimates 
of the total gas mass, which is incorporated within the halo gas. We 
approximate the extraplanar gas layer integrated over the galaxy disk
as a cylindric geometry which gives $dV = \pi\,|z|^2\,dR$, with z as the 
extraplanar radius, and $R$ as the radial extent of the galaxy. Under the 
assumption of $n_{\rm e} \approx n_{\rm p}$, and with $M = \int\,\rho\,dV$, 
and $\rho = n_{\rm e}\,m_{\rm p}$, this translates to 

\begin{equation}
M_{\rm DIG} = n_{\rm e}\,m_{\rm p}\,\pi\,|z|^2\,R
\end{equation}

The derived column densities, electron densities, and masses for our 
studied galaxies, which have diffuse extraplanar gas layers, are listed in 
Table~\ref{T5} below. 

For comparison we list the value of the diffuse gas mass of NGC\,891, 
including a correction for internal extinction, that was derived by Dettmar 
(\cite{De90}), which is $M_{\rm DIG} = 4\times 10^8\,M_{\sun}$. 

\begin{table}[h]
\setcounter{table}{4}
\caption[]{DIG properties of the galaxies with gaseous halos}
\label{T5}
\begin{flushleft}
\begin{minipage}{20cm}\small
\begin{tabular}{llll}
\noalign{\smallskip}
\hline
Galaxy & $N_{\rm H}[\rm{cm}^{-2}]$ & $n_{\rm e}[\rm{cm}^{-3}]$ & $M_{\rm DIG}
[M_{\sun}]$\\ \hline
NGC\,1963 & $1.05 \times 10^{19}$ & 0.0017 & $1.8\times10^6$\\
NGC\,3044 & $4.85 \times 10^{19}$ & 0.0054 & $1.3\times10^7$\\
NGC\,4302 & $6.05 \times 10^{18}$ & 0.0023 & $3.2\times10^6$\\
NGC\,4402 & $1.19 \times 10^{19}$ & 0.0012 & $5.0\times10^6$\\
NGC\,4634 & $4.36 \times 10^{19}$ & 0.0064 & $5.1\times10^6$\\
\noalign{\smallskip}
\hline
\end{tabular}
\end{minipage}
\end{flushleft}
\end{table}


\section{Discussion}

After the first detection of eDIG in external galaxies (e.g.,\, Dettmar 
\cite{De90}; Rand et al. \cite{Ra90}) the question was raised, whether it is 
a general case, that all late--type galaxies show DHI, or whether this is an 
exception. During the last decade a few investigations have been performed 
using basically optical narrowband imaging (H$\alpha$) in selected edge--on 
spirals. The sample sizes of these small--surveys were typically $\leq 10$ 
galaxies. (e.g.,\,Rand \cite{Ra96}; Pildis et al. \cite{Pi94}; Hoopes et al. 
\cite{HoWaRa}; this work). Additional studies in the optical regime have been 
carried out -- mostly irregular and dwarf galaxies (e.g.,\,Martin 
\cite{Ma97}) -- and on single objects such as \object{NGC\,55} (Ferguson et 
al. \cite{FeWy}), and three Sculptor group galaxies including also NGC\,55 
(Hoopes et al. \cite{HoWa}), among a few other galaxies. In the next section 
we discuss our observations and compare our results with other observations, 
such as the Lehnert \& Heckman starburst sample.

\subsection{Comparison between starburst and normal (quiescent) galaxies}

In an investigation by Lehnert \& Heckman (\cite{LeHe}) all IRAS bright 
and IRAS warm nearby edge--on starburst galaxies have been searched for 
extraplanar diffuse ionized gas. In this IR--selected survey 55 galaxies 
have been studied. This and supplementary studies have shown that all known 
nearby starburst galaxies show gaseous halos. These outflows most likely 
arise from starburst winds, driven by collective supernovae or massive star 
winds. This is most likely true for nuclear starbursts. However, presently it 
is not clear, if such a starburst wind is the driving force behind the 
outflows of normal or `quiescent' galaxies. In starburst galaxies the 
outflows occur preferentially from the nuclear regions (nuclear or central 
starburst), whereas in quiescent galaxies the filaments usually do not 
protrude from the nuclear region, rather from the strong SF regions 
distributed across the disk.

The criterium for IR--bright galaxies is quoted by $S_{60} \geq$ 5.4\,Jy and 
IR--warm galaxies are denoted by $S_{60}/S_{100} \geq 0.4$ (Lehnert \& 
Heckman \cite{LeHe}). Dahlem (\cite{Da97b}) lists also a value of $S_{60} 
\geq$ 30\,Jy for IR-warm galaxies. 

\begin{figure}[t]
\rotatebox{270}{\resizebox{6.0cm}{!}{\includegraphics{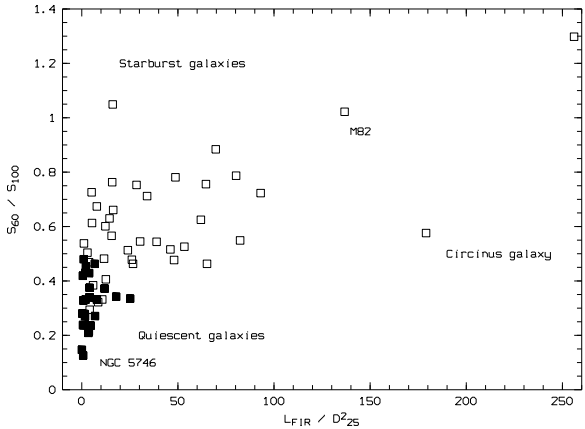}}}
\caption[]{Diagnostic diagram, showing the ratio of the flux densities 
at $60\mu$m and $100\mu$m expressed as $S_{60}/S_{100}$ versus the ratio 
of the FIR luminosity ($L_{\rm{FIR}}$) divided by the optical diameter of the 
$25^{\rm{th}}$\,mag/$\sq\arcsec$ isophote squared ($D^2_{25}$)) in units of 
$\rm{10^{40}\,erg\,s^{-1}\,kpc^{-2}}$. The open squares denote positions 
occupied by starburst galaxies whereas the filled squares denote locations of 
normal or `quiescent' galaxies. The starburst galaxies are essentially the 
sample studied by Lehnert \& Heckman (\cite{LeHe}). The galaxies indicated by 
the filled squares are galaxies currently investigated in the DIG context by 
various research groups (e.g.,\,Pildis et al. \cite{Pi94}; Rand \cite{Ra96}; 
this study).}
\label{F11}
\end{figure}

\begin{figure}[b]
\rotatebox{270}{\resizebox{6.0cm}{!}{\includegraphics{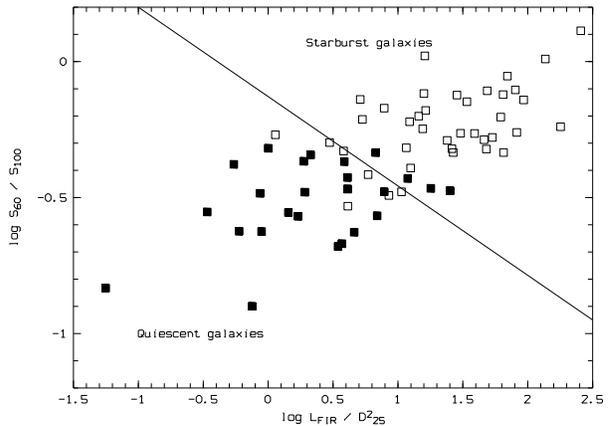}}}
\caption[]{Logarithmic diagram, showing the ratio of the flux densities 
at $60\mu$m and $100\mu$m expressed as $S_{60}/S_{100}$ versus the ratio 
of the FIR luminosity ($L_{\rm{FIR}}$) divided by the optical diameter of the 
$25^{\rm{th}}$\,mag/$\sq\arcsec$ isophote squared ($D^2_{25}$) in units 
of $\rm{10^{40}\,erg\,s^{-1}\,kpc^{-2}}$. The logarithmic plot shows 
the overlapping region in more detail. The open squares denotes areas 
occupied by starburst galaxies whereas the filled squares denote areas of 
normal or `quiescent' galaxies. The solid line separates the two areas 
occupied by the various galaxy types, with an overlapping region, since some 
normal galaxies are also listed as starbursts in the literature and vice 
versa.}
\label{F12}
\end{figure}

To illustrate differences between starburst galaxies and normal (quiescent) 
galaxies we have constructed diagnostic diagrams (see Fig.~\ref{F11}, and 
Fig.~\ref{F12}), in which the different types of galaxies populate different 
positions. We have plotted the ratio of the $60\mu$m and $100\mu$m IRAS fluxes 
as a function of the ratio of $L_{\rm{FIR}} / D^2_{25}$, which is the 
FIR--luminosity over the optical galaxy diameter of the $25^{\rm{th}}$ 
magnitude isophote squared. This term ($L_{\rm{FIR}} / D^2_{25}$) has been 
introduced by Rand (\cite{Ra96}) as a tracer for star formation activity in a 
galaxy (star formation rate per unit area).  

In Fig.~\ref{F12} we have plotted the same information in logarithmic scale, 
in order to show more detail on the border of starburst/non-starburst 
galaxies. The slope gives rather empirically a division between the two 
object classes. In the literature some galaxies which were initially 
identified as a starburst galaxy later were classified as a non starburst 
galaxy and vice versa. 

We have computed the FIR-luminosity according to 

\begin{equation}
L_{\rm{FIR}} = 3.1 \times 10^{39}\,D^2\,[2.58\,S_{\nu}(60) + 
S_{\nu}(100)]
\end{equation}

\noindent with $S_{\nu}$\,(60), $S_{\nu}$\,(100) the flux densities in Jy and 
with $D$ the distance to the galaxy in Mpc. The flux densities have been taken 
from the catalogue of Fullmer \& Lonsdale (\cite{Lo89}). The term $D^2_{25}$ 
is expressed in kpc$^2$.

The sample was compiled from the starburst galaxies studied by Lehnert \& 
Heckman (\cite{LeHe}), the normal (quiescent) galaxies were compiled from 
various investigations (e.g.,\,Pildis et al. \cite{Pi94}; Rand \cite{Ra96}; 
this work). In the following table (Table~\ref{T6}) we list some properties 
concerning the DIG for the non--starburst galaxies that were investigated in 
this study. 

\begin{table*}
\setcounter{table}{5}
\caption[]{Properties for diagnostic DIG diagrams}
\label{T6}
\begin{flushleft}
\begin{minipage}{20cm}\small
\begin{tabular}{lllllllll}
\noalign{\smallskip}
\hline
Galaxy & Type & T\footnote{mean numerical index of stage along the Hubble 
sequence in RC2 system} & $D$\,[Mpc] & $D_{25}$\,[$\arcmin$] & 
$S_{60}/S_{100}$ & $\rm{L_{FIR}\,[10^{43}\,erg\,s^{-1}]}$ & 
$\rm{L/D^{2}_{25}\,[10^{40}\,\frac{erg}{s\,kpc^2}]}$ & extraplanar dust feat. 
\\ \hline
NGC\,4634 & Sc & 6.0 & 19.1 & 2.57 & 0.3720 & 2.4247 & 11.892 & yes \\ 
NGC\,4402 & Sb & 3.0 & 22.0 & 3.89 & 0.3320 & 4.8664 & 7.853 & yes \\ 
NGC\,1963 & Sc & 5.5 & 17.7 & 2.82 & 0.4051 & 1.4494 & 6.954 & yes \\ 
NGC\,3044 & SBb & 5.0 & 17.2 & 4.90 & 0.4632 & 4.0546 & 6.746 & \\ 
NGC\,4302 & Sc & 5.0 & 18.8 & 5.50 & 0.2136 & 3.3496 & 3.703 & yes \\ 
UGC\,10288 &Sc & 5.3 & 27.3 & 4.79 & 0.2370 & 1.2882 & 0.890 & \\
NGC\,5170 & Sc & 5.0 & 20.0 & 8.32 & 0.2796 & 0.7940 & 0.339 & no \\ 
IC\,4351 & Sb & 3.0 & 35.5 & 5.75 & 0.1947& 5.7577 & 1.633 & no \\
IC\,2531\footnote{IC\,2531 has not been detected with IRAS. Therefore 
no information concerning the FIR properties is available} & Sc & 5.3 & 33.0 
& 6.92 & n & n & n & yes \\ 
\noalign{\smallskip}
\hline
\end{tabular}
\end{minipage}
\end{flushleft}
\end{table*}

The optical galaxy diameters $D_{25}$ have been taken from the RC3 (de 
Vaucouleurs et al. \cite{Vau91}). Extraplanar dust detections are based on 
a study of our R--band images. This investigation will be described 
separately in more detail (Rossa \& Dettmar, in prep.).

In the following figure (Fig.~\ref{F13}) we have constructed a diagnostic DIG 
diagram (DDD) for our studied galaxies. Galaxies with extraplanar gas layers 
have been denoted by filled squares whereas galaxies with no extraplanar gas 
are indicated with open squares. The open triangle denotes the galaxy with 
extraplanar gas features (e.g.,\,plumes), and where a weak extended layer has 
been detected. The plot is shown in a logarithmic scale.

\begin{figure}[h]
\rotatebox{270}{\resizebox{6.0cm}{!}{\includegraphics{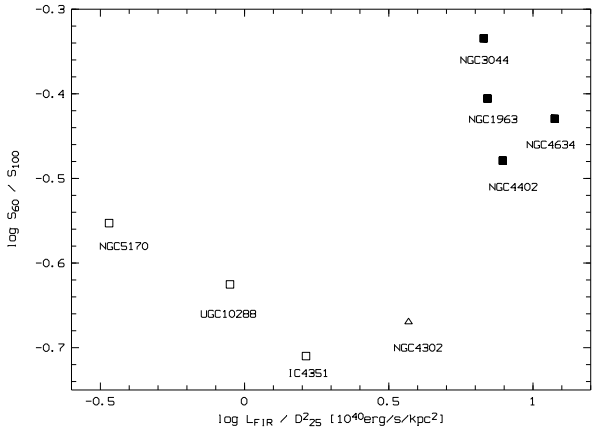}}}
\caption[]{Logarithmic diagram, showing the ratio of the flux densities 
at $60\mu$m and $100\mu$m expressed as $S_{60}/S_{100}$ versus the ratio 
of the FIR luminosity ($L_{\rm{FIR}}$) divided by the optical diameter of the 
$25^{\rm{th}}$\,mag/$\sq\arcsec$ isophote squared ($D^2_{25}$) in units 
of $\rm{10^{40}\,erg\,s^{-1}\,kpc^{-2}}$. The labels for the galaxies of our 
studied sample have been indicated. Filled squares denotes detections of 
extraplanar DIG layers, open squares indicate those with no detections, 
whereas the open triangle denotes the galaxy with eDIG features, but where 
only a weak layer has been detected.}
\label{F13}
\end{figure}

Even with such a small sample it is evident that the galaxies with eDIG layers 
occupy certain regions in the DDD, whereas galaxies without eDIG halos do 
appear in different locations in this diagram. It is also clear that a larger 
sample of galaxies is necessary to make a quantitative approach in order to 
find a limiting value of the ratio of $\frac{S_{60}}{S_{100}} / 
\frac{L_{\rm{FIR}}}{D^2_{25}}$ above that extended gaseous halos do exist. 

Investigations to find this lower threshold are currently underway (Rossa \& 
Dettmar, in prep.), where a much larger, distance limited and unbiased (in 
the sense of star formation activity) sample is covered. 
   
\subsection{Energetics of the outflow condition}

In a study by Dahlem et al. (\cite{DaLi}) it was concluded that the energy 
input of star formation regions into the ambient ISM, powered by SNe, is 
largest for galaxies with large and extended radio halos. In cases of low or 
non--detectable outflows (both H$\alpha$ and radio continuum) such as 
NGC\,4244 (Hummel et al. \cite{HuSa}), the faint end of the energy input rate 
has been reached. They use an important parameter as a measure of the 
disk--halo interaction (DHI), which is the threshold value of the energy 
injection into the ISM, which is given by $\frac{dE^{\rm{tot}}_{\rm{SN}}}{dt}/
A_{\rm{SF}} = \nu_{\rm{SN}}\,E_{\rm{SN}}/2\pi\,r^2_{\rm{SF}}$. Here the total 
energy injection is the input from SNe ($\rm{E_{SN}}=10^{51}\,\rm{erg\,s^{
-1}}$). Depending on the derived SN rates ($\nu_{\rm{SN}}$), which might vary 
from 0.01 yr$^{-1}$ to 0.05 yr$^{-1}$ (van den Bergh \cite{vdB91}, and Dahlem 
et al. \cite{DaLi}, respectively), mean values of the total energy input of 
$1.5\times10^{-4}\,\rm{erg\,s^{-1}\,cm^{-2}}$ are derived. 

However, gravitational interactions between the galaxies, such as in the case 
of NGC\,4631 can lead to deviations from this trend that only high energy 
inputs powered by SNe lead to strong outflows. Therefore isolated galaxies 
should be considered as suitable candidates. The energy injection is of 
course highly depending on the extend of the star formation activity area, 
where strong local SF activity is also in favor of strong outflows. This 
should be taken into account when selecting candidate galaxies with star 
formation driven outflows. Therefore a bias might result, if only FIR bright 
galaxies are selected. Galaxies with high local SF, despite a low FIR flux, 
would not be covered by this selection criterium. This is an important issue 
for future investigations.    


\section{Summary}

We have presented the results of a small H$\alpha$ survey aiming at the 
detection of extraplanar diffuse ionized gas (eDIG). 9 galaxies have been 
investigated. 4 of them show eDIG, whereby 2 others show only plumes, 
filaments, and in the case of NGC\,4302 a faint eDIG layer is present. 3 
galaxies do not show any signs of disk--halo interaction. The star formation 
rates have been derived for those edge--on galaxies, which showed bright 
diffuse emission. The morphology of the bright eDIG in NGC\,4634 is similar 
to that of previously studied galaxies (e.g.,\,NGC\,891 and NGC\,5775), 
although the emission is not as bright and far extended as in NGC\,891. 
Furthermore a difference in comparison to NGC\,891 is that the eDIG layer in 
NGC\,4634 is symmetrically and homogeneously distributed, unlike the 
asymmetry which is observed in NGC\,891. 

The extraplanar filaments in NGC\,4634 may be compared with the theoretical 
models such as the `chimney' scenario by Norman \& Ikeuchi, where the 
individual filaments (seen also in several other edge--on galaxies) may 
represent these very {\em chimneys}. However, this could probably only be 
investigated with high angular resolution studies, such as HST WFPC\,2 
observations. We have presented new evidence for disk--halo interaction in 
late--type galaxies (Sb--Sc). The fraction of DIG in the halo has been 
derived, and show that a significant mass of the gas is present in galaxy 
halos. It has become clear that eDIG is not observed in all late--type 
galaxies, and we argue that the DHI seems not to be a general case for all 
normal spirals. eDIG seems to be correlated with the star formation activity 
in the underlying disk, which is evidenced by the presence of star forming 
regions (e.g.,\,\ion{H}{ii} regions) below the filaments. As a consequence, 
the galaxies with low SF activity show no or almost no gas outflows at the 
detection limit, which is given by the instrumentation. It also reflects the 
cosmic evolution of spiral galaxies with the episodes of star formation. The 
DIG mass has been derived for galaxies with extended DIG. These masses 
represent lower limits, since no correction for internal extinction has been 
applied. 


\begin{acknowledgements}
The authors would like to thank Deutsches Zentrum f\"ur Luft-- und 
Raumfahrt (DLR) for financial support of this research project through 
grant 50\,0R\,9707. It is a pleasure to thank Dr.~D.~Bomans for his
helpful comments on the manuscript. We would like to thank the anonymous 
referee for useful suggestions, which helped to improve the presentation 
of the paper. This research has made extensive use of the NASA/IPAC 
Extragalactic Database (NED) which is operated by the Jet Propulsion 
Laboratory, California Institute of Technology, under contract with the 
National Aeronautics and Space Administration.      
\end{acknowledgements}



\end{document}